# Chaos in some young asteroid families

Rosaev A., Plavalova E

Research and Educational Center of Nonlinear Dynamics, Yaroslavl State University, Yaroslavl, Russia
Astronomical Institute Slovak Academy of Sciences

**Abstracts**

Asteroid families are groups of minor planets that have a common origin in breakup events. The very young compact asteroid clusters are the natural laboratory to study resonance related chaotic and nonlinear dynamics.

The present dynamical configurations and evolution of asteroid associations strongly depends on their ages. In present paper we allocate subclass of very young asteroid families (younger than 1 Myr).

We show that resonance-related chaos can play a very important role in dynamics of very young asteroid families. In case of Datura family chaos may be explained by high order mean motion resonance 9:16 with Mars. In case Hobson family chaos is affected by secular resonance. In other considered cases (Kap'bos cluster and Lucascavin cluster) origin of chaotic behavior is still unknown. The effect of resonance is very selective in all cases: we see very stable orbits in the vicinity of chaotic ones. In the high order resonance transfer from initial to final orbit take place by temporary capture in exact resonance.

The large asteroids (Ceres, Vesta) can made significant effect on dynamic of small bodies in resonance. In some cases (as for Datura and Lucascavin family), their perturbations can extend area of chaotic motion.

## Introduction

Hirayama (1918) was the first to detect concentrations of asteroids with very similar orbital elements, now known as asteroid families. These groups of minor planets are believed to have originated from catastrophic breakups of single parent bodies. Now asteroid families are intensively investigated because they provide us with unique possibilities of studying different processes related to high-energy collisions. At the present time, more than 120 families have been discovered across the asteroid main belt. A summary of the current state of affairs in the field of family identification is given in Bendjoya & Zappala (2002)**[1];** Carruba & Michtchenko (2007); Carruba et al. (2013)**[2,3].**

Throughout the article we have used standard notations for orbital elements *a* – semimajor axis in a.u., e – eccentricity, *i* – inclination, $\Omega$ – longitude of ascending node, ω – perihelion argument, $\varpi$ – longitude of perihelion (the angular elements are in degrees).

To study the dynamical evolution of young asteroid clusters, the equations of the motion of the systems were numerically integrated 800 kyrs into the past, using the N-body integrator Mercury **(Chambers, 1999)[4]** and the Everhart integration method **(Everhart, 1985)[5]**.

Asteroid families traditionally classified into old and young (younger 1Gyr). Most of families are old (older 1 Gyr). The discovery of very young asteroid clusters (e.g., Nesvorný et al. 2006 [7]; Nesvorný & Vokrouhlický 2006[6]; Pravec &Vokrouhlický 2009 [8]) open a new phase in the analysis of the asteroid families. In very young families we have become aware of many dynamical and physical processes that modify **the orbital evolution** of their individual members with time. In very young families, of ages less than one million years, most of these processes had insufficient time to operate.     From their study, we might be able **to reconstruct** properties of their parent object fragmentation **and to determine their age with more precision** [7]

Here we argue in favor to outline very young (with age smaller than 1Myr) family subclass. At this moment 10 very young asteroid clusters are known.

There is remarkable difference between young and very young asteroid families, based on YORP effect action. In young families the YORP-driven variation of semimajor axis is about $\Delta = 10^{-4} - 10^{-5}$ a.u. It is comparable with dispersion da of semimajor axis in family, so we can count, that large part of da is due to YORP effect. But it is not true for very young families where $\Delta$ is by order smaller. It means, that significant part of da in the very young family produced at the epoch of breakup immediately.

Young and very young families are different by number of members. By this reason they will be different by method of studying. In addition, youngest families are more compact in angular orbital elements.

There are few other factors, over the Yarkovsky thermal forces, which can produce small divergence in orbital elements and mutual distances at the moment of impact. **Following Nesvorny and Bottke [9] we neglect in youngest asteroid families:** (iii) the direct effect of the Yarkovsky force on the apsidal and nodal rates is negligible in the current context; (iv) the integration errors; (vi) gravitational perturbations by Mercury have negligible differential effects on $\delta\Omega_P$ and $\delta\varpi_P$; (viii) the differential relativistic effects on $\delta\varpi_P$ are negligible.

Commonly, youngest families have small number of members. The important characteristic of youngest families is their compactness in angular variables $\Omega$ and $\varpi$. In contrary, young Karin cluster have a wide range for these elements. The possible explanation of this difference is subsequent collisions in young families, which are absent in youngest ones.

But some factors, not important for young (1Gyr<age<1Myr) families, may be significant at studying youngest clusters. In particular, we show in present paper that chaos effect and Ceres perturbations may be significant for some very young asteroid families

In [9] is claimed that the chaos influencing asteroid orbits at the location of the Karin young cluster is negligible on <1 Myr timescales. Here we show, that resonance related chaos may be important in case Datura and Hobson young families.

In [9] suggested following expressions for angles differences:

$$\Omega_j = \Omega_{j0} + s\tau + \frac{1}{2}\frac{\partial s}{\partial a}\dot{a}_j\tau^2$$

$$\varpi_j = \varpi_{j0} + g\tau + \frac{1}{2}\frac{\partial g}{\partial a}\dot{a}_j\tau^2$$

**Here** $\Omega_{j0}$ and $\varpi_{j0}$ – initial values of orbital elements of j-th member of cluster at the moment of breakup, g and s – proper frequencies, $\tau$ is an age of family. Evidently, we can write the similar equation for perihelion argument:

$$\omega_j = \omega_{j0} + l\tau + \frac{1}{2}\frac{\partial l}{\partial a}\dot{a}_j\tau^2$$

**We build the dependences of $\delta\Omega$ and $\delta\omega$ on $\tau$ for known very young asteroid families with age < 1Myr (fig. 1, 2). These dependences are very close to linear. This remarkable fact has two important followings. First of all, YORP-related semimajor axis drift is not important in very young families. Secondly, the previous ages estimations are close to true values.**

**Moreover, in supposition that orbital elements of all members of family are equal at the moment of breakup, we can try to exclude semimajor axis drift in equations above and estimate age of family:**

$$\tau = \left[\frac{\partial g}{\partial a}\delta\Omega_{ij} - \frac{\partial s}{\partial a}\delta\varpi_{ij}\right] / \left[\frac{\partial g}{\partial a}\delta s_{ij} - \frac{\partial s}{\partial a}\delta g_{ij}\right]$$

**Here** $\delta s = s_i - s_j$ and $\delta g = g_i - g_j$ are function of time.

At this moment, this expression is only formal illustration how YORP effect can be reduced at the age of family estimation. For the practical application, in particular, this method required the high precision of proper elements computations.

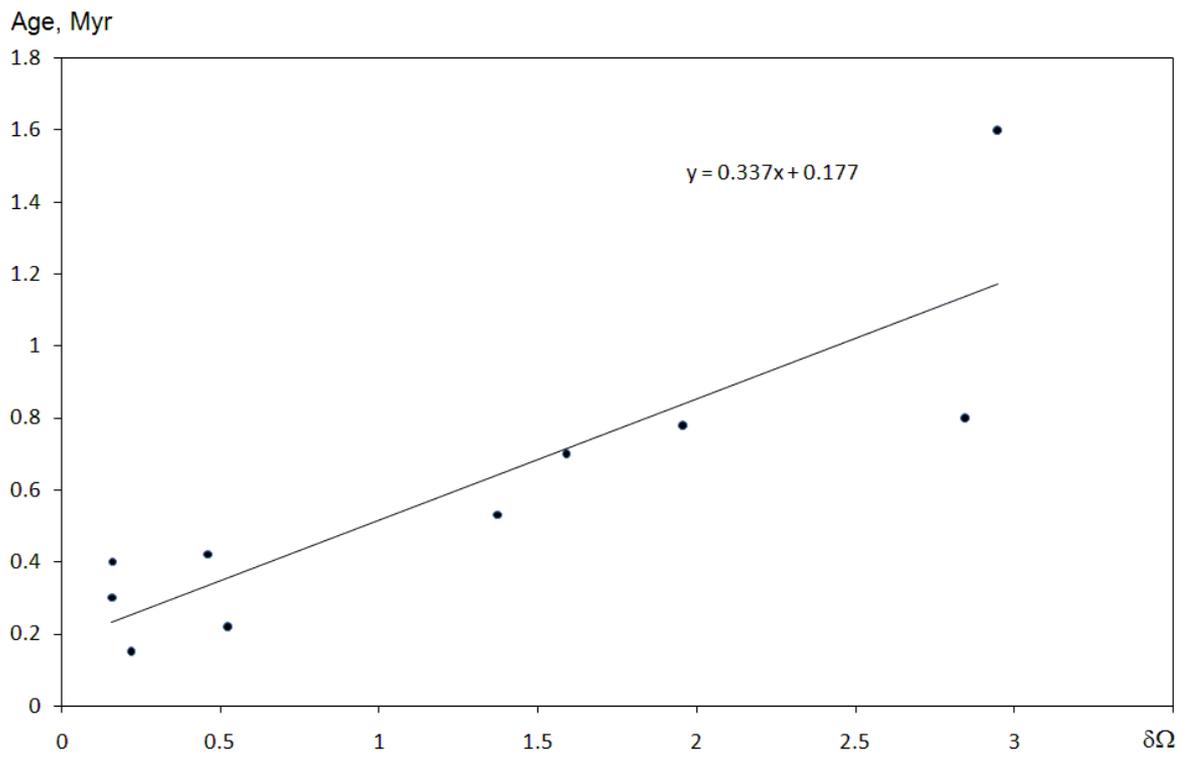

Fig.1. The dependence ages of very young families on node longitude dispersion

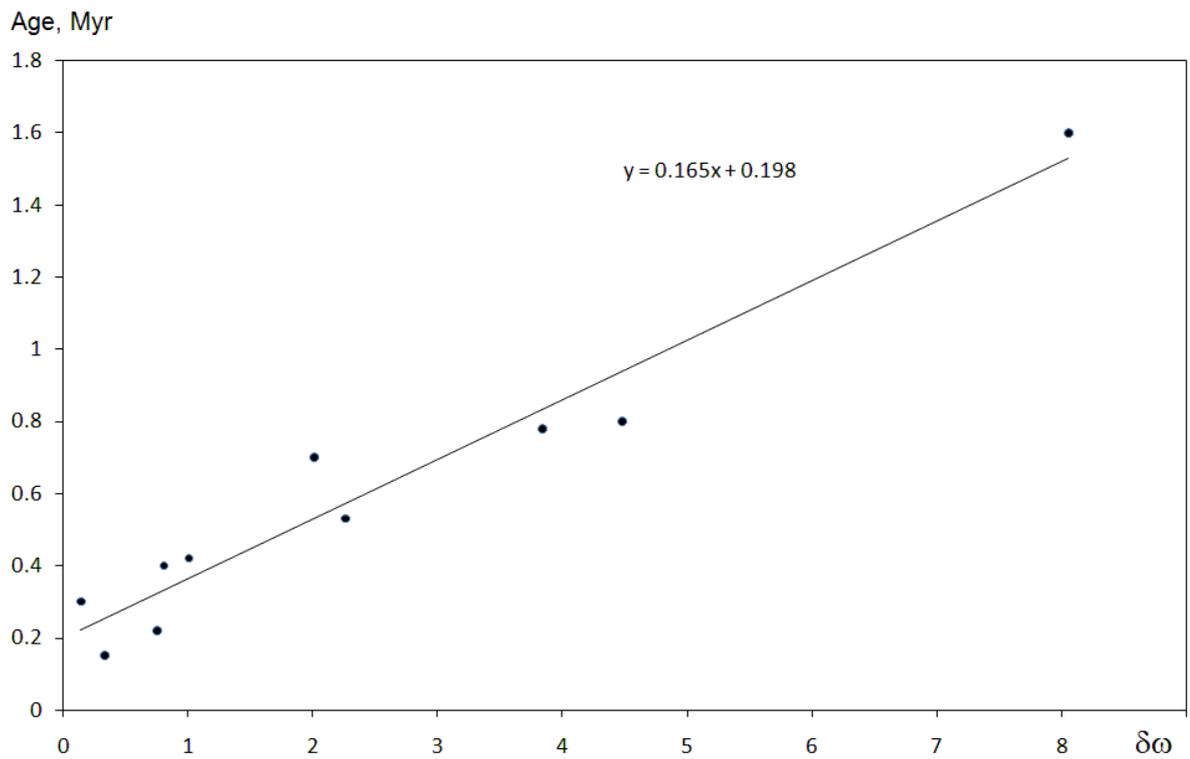

Fig.2. The dependence ages of very young families on perihelion argument dispersion

# Chaos in Datura family

A group of minor planets associated with the largest body Datura(1270), is of particular interest because it has enough known members and resides in the inner part of the main asteroid belt, which makes observation clearer. The Datura family was discovered by Nesvorny et al. (2006) [6], after which, both its dynamical evolution and physical properties were rigorously studied. The Datura family consist of one large 10 km-sized asteroid (parent body) and a few small minor planets; possible fragments of a catastrophic breakup. Up to recent time only 7 members of the Datura family were known (Nesvorny et al., 2015, 2006 [6,10]; Vokrouhlicky et al., 2009[11]), all of them belonging to the so-called S-type of minor planets Tholen (1984) [12]. In our previous paper (Rosaev, Plavalova, 2015) [13] we report about 3 new members (table 1).

Table 1: Osculating orbital elements of the Datura family members at epoch 16-01-2009 (JD 2454848).

| Object | | $\omega$ | $\Omega$ | $i$ | $e$ | $a$ |
|---|---|---|---|---|---|---|
| 1270 | Datura | 258.84072020 | 97.88191931 | 5.98954802 | 0.207917356 | 2.234316251 |
| 60151 | 1999 UZ6 | 260.57151084 | 96.79987218 | 5.99341398 | 0.207812911 | 2.235186383 |
| 89309 | 2001 VN36 | 266.85629891 | 92.976202113 | 6.02164403 | 0.20634112 | 2.23559504 |
| 90265 | 2003 CL5 | 261.84375257 | 95.69777810 | 5.99543662 | 0.207471114 | 2.234865786 |
| 203370 | 2001 WY35 | 260.44606900 | 96.87210762 | 5.99154981 | 0.207435802 | 2.235226671 |
| 215619 | 2003 SQ168 | 259.39568050 | 97.46790499 | 5.99082425 | 0.207908027 | 2.234266272 |
| **338309** | **2002 VR17** | 260.60046232 | 96.80636247 | 5.98940220 | 0.207747446 | 2.235169351 |
| | 2003 UD112 | 263.12446374 | 95.47877895 | 6.00471739 | 0.206934529 | 2.234566113 |
| | **2002 RH291** | 262.04302362 | 95.75928088 | 5.99579553 | 0.207613218 | 2.235326612 |
| | **2014 OE206** | 261.72701226 | 96.26530222 | 6.00041657 | 0.206975298 | 2.235606600 |

NOTE: The three new members of the Datura family are shown in bold.

The orbit of 2014 OE206 is most interesting in context resonance related chaotic dynamic. As it was adopted by Nesvorný et al. [6], the orbit of (89309) 2001 VN36 was assumed to be too uncertain because of resonance-related chaoticity. (mean motion resonance 9:16 with Mars[7]). Except noted above, there is commensurability 7:2 with Jupiter in vicinity of Datura family but its effect is more weak. Initial mean anomalies and mean motions of (89309) 2001 VN36 and 2014 OE206 are close, so we can expect resonance perturbation of its orbit. Really, how followed from our results, semimajor axis of 2014 OE206 significantly increase at time about 50 kyr ago (fig.3). Elementary consideration is show, that there is not encounter with Mars close 0.01 a.u. In fact, orbit 2014 OE206 is a typical example of so-called stable chaotic orbit (Milani et al., 1997) [14]. The change of semi-major axis in the case of 2014 OE206 is significant; 50 kyrs ago, this asteroid leaped from one side of 9 : 16 resonance to another. On the other hand, its other orbital elements remained very close to the mean elements of the other members of the Datura family.

As a possible reason to catastrophic jumping about 50 kyr ago may be Mars orbit variation. It is evident, that position of 9:16 M resonance is not constant in time due to Mars orbit evolution. Detail view of orbit 2014 OE206 just before and after semimajor jumping shows typical resonance behavior (fig. 4). We can claim that so-called stable chaotic behavior provided through temporary capture in resonance.

Integration of orbit of 2014 OE206 and some clones show the following (Fig.5). This effect presents at variation initial semimajor axis of this asteroid in range 2.23485<$a$<2.23505 a.u., and initial mean anomaly in range at least 36.70<M<36.90. The variations of initial values of other orbital elements (eccentricity, inclination, pericenter and node longitudes) have relative small effect on 2014 OE206 orbital evolution. The core (central zone) of resonance contains regular orbits flanked both sides by chaotic zones. Orbits in these zones show so-called regular chaotic behavior jumping from one side of resonance to another. Jumps take place in both directions: from inside resonance to outside and by the opposite way. The loss of stability is by temporary capture in resonance. The amplitude of resonance-related bifurcations (in *a*) decreases close to core of resonance, when frequency of bifurcations is increases. The described dynamical behavior may be understands in the context of the model of the mean motion resonance multiplicity [15] (or splitting [16]).

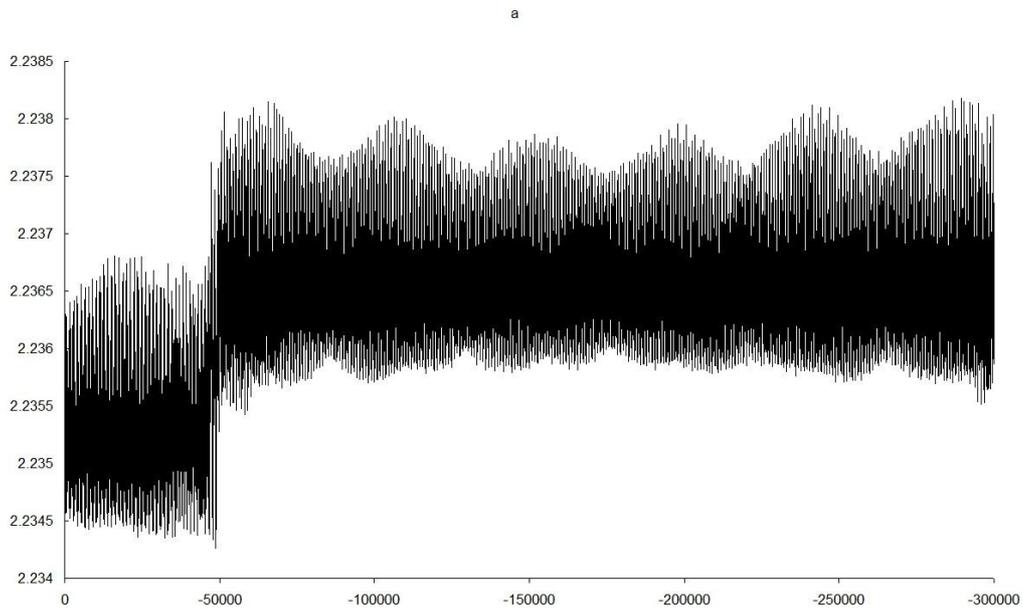

Fig.3. The evolution of the semi-major axis of 2014 OE206 close to -50 kyrs epoch.

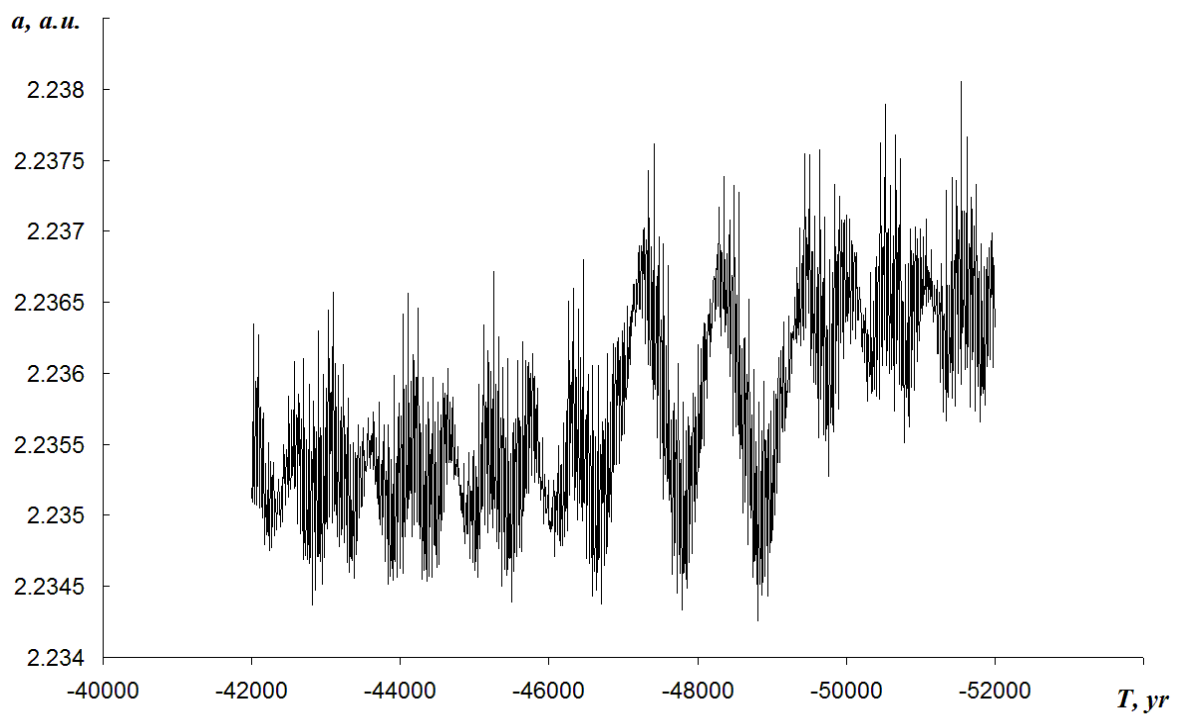

Fig.4. The same as in fig.1 but in more detail scale

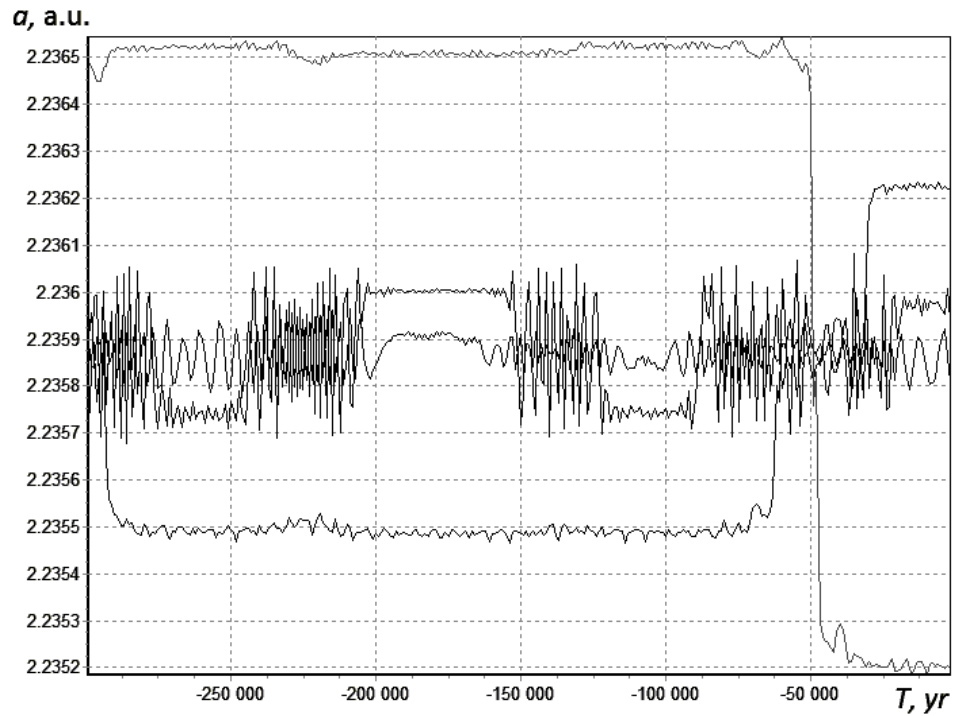

Fig. 5. The evolution of the semimajor axis of orbits of clones in vicinity 9:16 mean motion resonance with Mars

In the process of orbital evolution, zone of 9:16 resonance has shifted due to Mars orbit variation. Close to epoch 50 kyrs ago, effect on dynamics of Datura family members become maximal.

**The Ceres and Vesta perturbations of Datura family**

As in previous papers devoted to the Datura family, all the above results were obtained by the integration of large planets only. But as (Carruba et al., 2016; Novakovic et al., 2015) **[17,18]** rightly noted, large asteroids (Ceres, Vesta, Juno, Pallas) can make a remarkable impact on the perturbations of the main belt asteroids. For this reason, we repeated our numerical integration with the inclusion of Ceres and Vesta as additional massive perturbers. Using this new data, our main results have significantly changed. First of all, in the presence of the Ceres and Vesta perturbations, we observe a behaviour similar to that of 2014 OE in the semi-major axis of other members of the Datura family. The Ceres and Vesta perturbations triggered an effect which destabilized the orbit of 89309 (2001 VN36) (Fig.6). Secondly, the convergence of angular elements for this asteroid relative to 1270 Datura changed. We can conclude that effect of Ceres on the dynamics of the Datura family is quite significant and cannot be neglected.

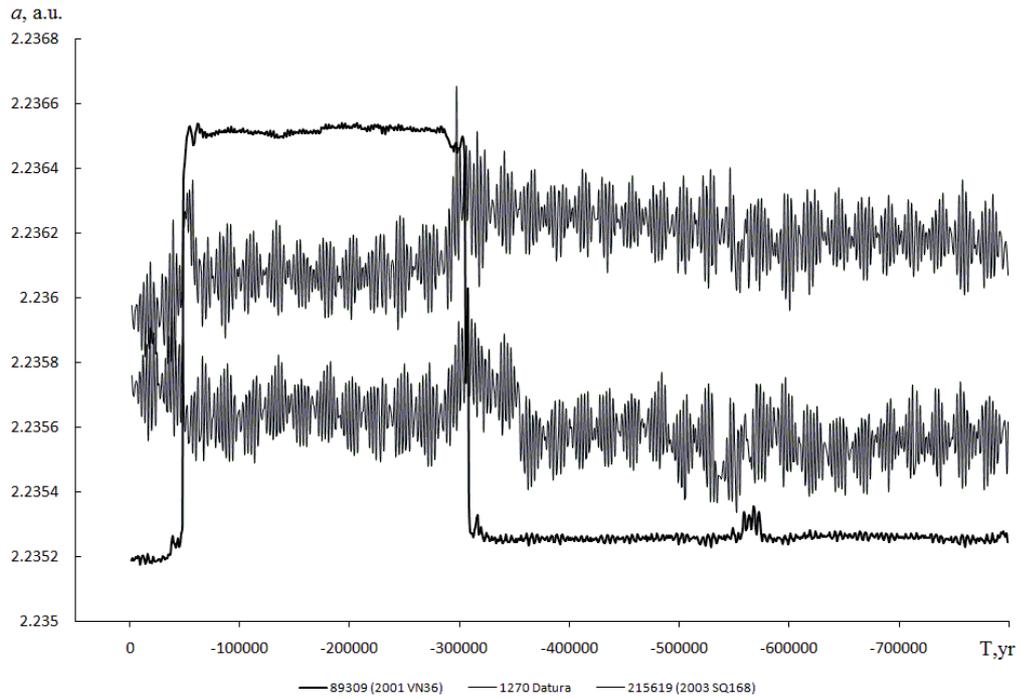

Fig. 6.

As it easy to see in fig 6, it is not effect of close encounters, because orbits of other members of family have remarkable perturbations at the same epoch. Definitely, it is not a low-order (1:1) mean motion resonance and not secular Ceres or Vesta resonance. The most natural explanation is the trigger effect, when Ceres perturbations move asteroid in zone of action 9:16 Mars mean motion resonance. The combination of these effects leads to instability and irregular changes in asteroid semimajor axis.

### The Hobson family: the chaotic orbit of 18777 Hobson

The cluster associated with asteroid 2001 UZ160(57738) was discovered by Pravec & Vokrouhlicky (2009) **[8]**, who noted its resonance perturbation (J3/1 mean motion resonance and g + g5 - 2g6 secular resonance). However, there are some problems which require a more detailed study; maybe for this reason, this cluster is not included in the list of young asteroid families (Nesvorny et al., 2015).

The original result of the search for Hobson members in close orbit in the asteroid belt made by Pravec & Vokrouhlicky (2009) **[8]**. They estimate age of Hobson cluster smaller than 500 kyrs. In our recent paper [19], we refine age estimation for this family (**-365 ±67 kyrs**) and add a new member **436620 (2011 LF12)** (table 2).

Table 2. Osculating orbital elements of 18777 Hobson family. Epoch 2451000.5 = A.D. 1998-Jul-06 00:00

| | Object | $\omega$ | $\Omega$ | $i$ | $e$ | $a$ |
|---|---|---|---|---|---|---|
| **18777** | Hobson | 178.75574 | 105.59056 | 4.3179967 | 0.1836943 | 2.56226765 |
| **57738** | 2001 UZ160 | 180.50332 | 105.03622 | 4.3269566 | 0.1842270 | 2.56217663 |
| **363118** | 2001 NH14 | 181.70784 | 105.34650 | 4.3168472 | 0.1817412 | 2.56365878 |
| **381414** | 2008 JK37 | 180.84799 | 104.40340 | 4.3313187 | 0.1839642 | 2.56231868 |
| **436620** | 2011 LF12 | 178.65699 | 105.09260 | 4.3139072 | 0.1813426 | 2.56312112 |

**Orbit of** 18777 Hobson becomes a very chaotic about 400 kyr ago. Two clones with initial difference in semimajor axis $\delta a = 0.00000008$ AU are rapidly diverge one from another (fig.7). The chaos appears in semimajor axis evolution: the difference between semimajor axis 18777 Hobson and mean semimajor axis of family has a significant trend between 200 and 400 kyrs ago (fig.8). This behavior is differ from a stable chaos related with high order resonances which described in [14].

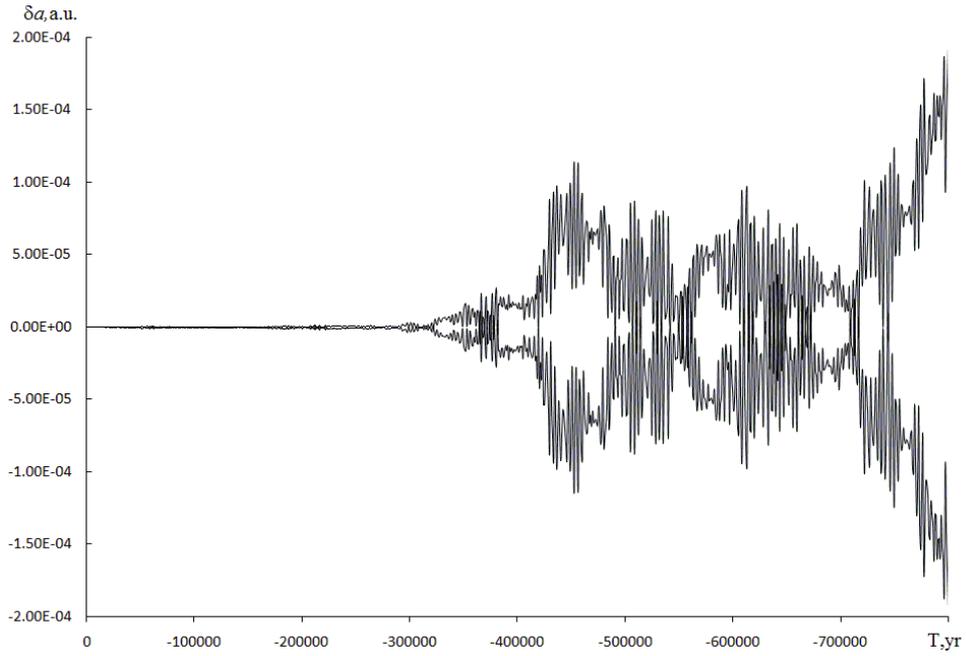

**Fig.7**.The difference in semimajor axis of two clones 18777 Hobson. Initial difference in a=8*10$^{-8}$ a.u.

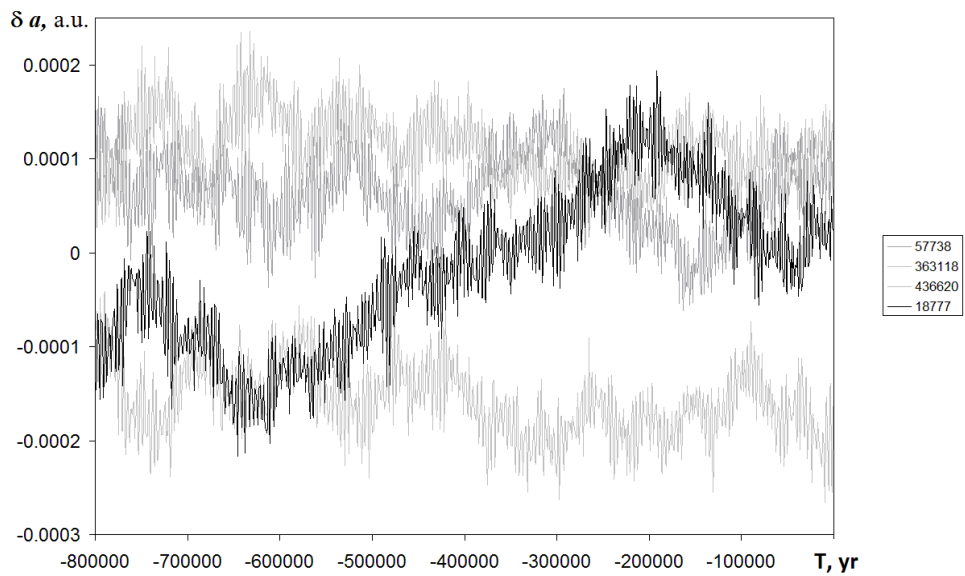

**Fig.8**.The difference in semimajor axis members of Hobson family and mean semimajor axis of cluster. Case 18777 Hobson marked black line.

In the close neighborhood of the orbit of 18777 Hobson we find a very regular orbit of **436620 (2011 LF12),** fig 9. It means that area of chaotic motion near the studied secular resonance is very narrow.

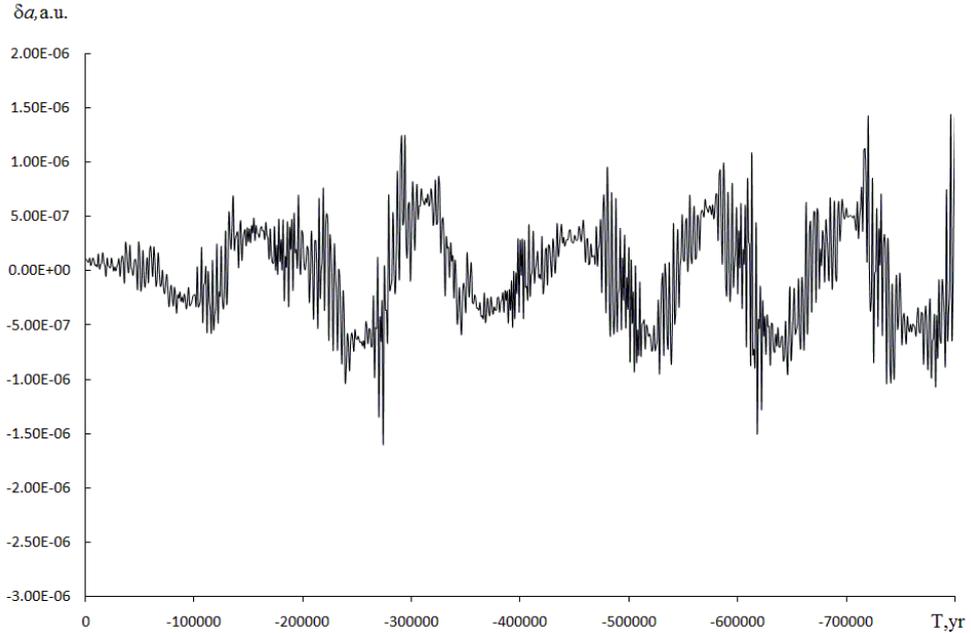

Fig.9. The difference in semimajor axis of two clones 2011 LF12 clones. Respect difference in initial semimajor axis $\Delta a$=-0.0000001 a.u.

There are minor changes after the Ceres and Vesta perturbations have taken into account. The orbit of 18777 Hobson show a significant variations of semimajor axis but their amplitude is rather smaller. The regular character of orbit 436620 (2011 LF12) have not changed when Ceres perturbation is take into account. In more detail, the effect of Ceres on Hobson family dynamic is described in [19].

### Kap'bos family

Other interesting example of chaotic behavior is a compact young family related with asteroid 11842 Kap'bos which is moved in inner part of the main asteroid belt. The inner zone of the asteroid main belt is dominated by the large and diffuse Flora family.

The Flora family resides in the densely populated inner main belt, bounded in semimajor axis by the v6 secular resonance and the Jupiter 3:1 mean motion resonance. The presence of several large families that overlap dynamically with the Floras (e.g., the Vesta, Baptistina, and Nysa-Polana families), and the removal of a significant fraction of Floras via the nearby v6 resonance complicates the Flora family's distinction **in the proper orbital elements.** [20]

Before present, the pair of close orbits 11842 Kap'bos and 228747(2002VH3) was known [8]. First notation about Kap'bos family present in [21].

Here we report about third asteroid 436415 (2011 AW46) with orbit close to this pair (table 3, 4), so pair becomes a compact family like Emilkowalski or Lucaschavin cluster. Note, that it is only one case when we can find and append a third body to pair. The largest body in this cluster is minor planet **11842 Kap'bos (1987 BR1)** have absolute magnitude H=13.9 mag. Other objects are smaller (16.7 and 18.2 mag).

Table 3. Osculating orbital elements of Kap'bos cluster

| asteroid | | $\omega$ | $\Omega$ | $i$ | $e$ | $a$ |
|---|---|---|---|---|---|---|
| 228747 | **2002 VH3** | 172.127724 | 273.099249 | 3.690776 | 0.09430342 | 2.25026200 |
| 11842 | **Kap'bos** | 172.346045 | 272.844049 | 3.688654 | 0.09440498 | 2.24992700 |
| **436415** | 2011 AW46 | 172.395639 | 272.810987 | 3.688923 | 0.09442976 | 2.25007100 |

Table 4. Proper orbital elements of Kap'bos cluster [22]

| asteroid | | g | n | Sin(*i*) | *e* | *a* |
|---|---|---|---|---|---|---|
| 228747 | **2002 VH3** | 34.1138 | 106.632 | 0.074670 | 0.12132 | 2.25033 |
| 11842 | **Kap'bos** | 34.1130 | 106.633 | 0.074678 | 0.12134 | 2.25031 |
| **436415** | 2011 AW46 | 34.1184 | 106.621 | 0.074681 | 0.12128 | 2.25047 |

Results of our first integration, using only large planets perturbations show a very regular evolution of semimajor axis of members of family (black line on fig.8). But when we have taken into account Ceres and Vesta perturbations, we obtain a chaotic evolution with multiple irregular changes of 228747(2002VH3) semimajor axis (blue line on fig. 10). The orbit of **228747**(2002VH3) is most perturbed by large asteroids. The chaotic character of orbit 228747(2002VH3) has not changed at small variation of initial elements of its orbit. In the same time, orbits of two other members of cluster stay to be regular with small short-periodic variations of semimajor axis.

Finally, we can state following. Large asteroids (Ceres, Vesta) can a very significant perturbations of orbits of very young compact asteroid families, in some cases principally change the stability of motion.

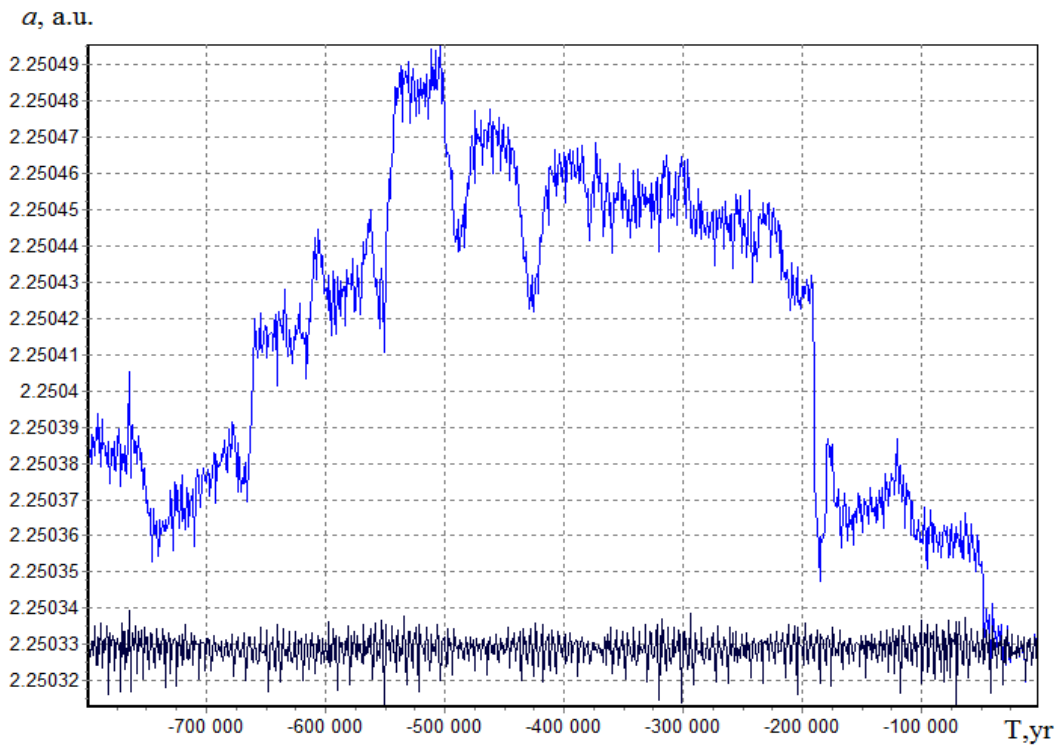

**Fig.10.** Chaotic evolution of 228747(2002VH3) semmajor axis with Cerees Vesta and Pallas perturbations (blue line) and with large planets perturbations only (black line)

Noted chaotic behavior significantly complicates the estimation of the age of cluster. The convergence of angular elements is very poor and we can note only wide range of possible age: 0-330 kyrs. Some addition information we obtain in mutual encounters. In the first our integration (without large asteroids perturbations) the evolution of the mutual distances of asteroids in each pairs confirms a very long and stable period between subsequent encounters in each pair. After the Ceres and Vesta perturbations have taken into account, orbit of 228747(2002VH3) become chaotic, but mutual encounters between 11842 Kap'bos and 436415 (2011 AW46) still rare (fig. 11-13). Based on this, we can note epoch around -240 kyrs as a possible origin of Kap'bos cluster.

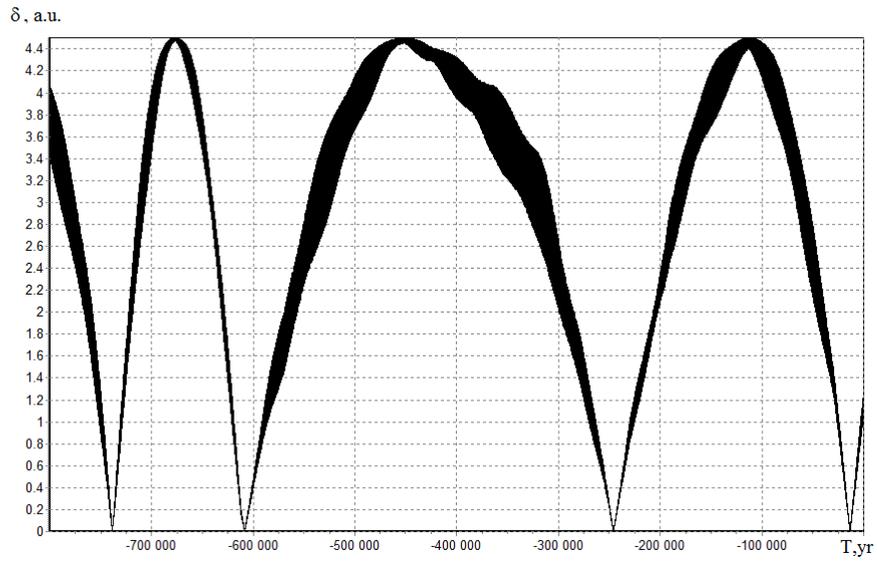

**Fig.11** Evolution distance between 11842 Kap'bos and 436415 (2011 AW46)

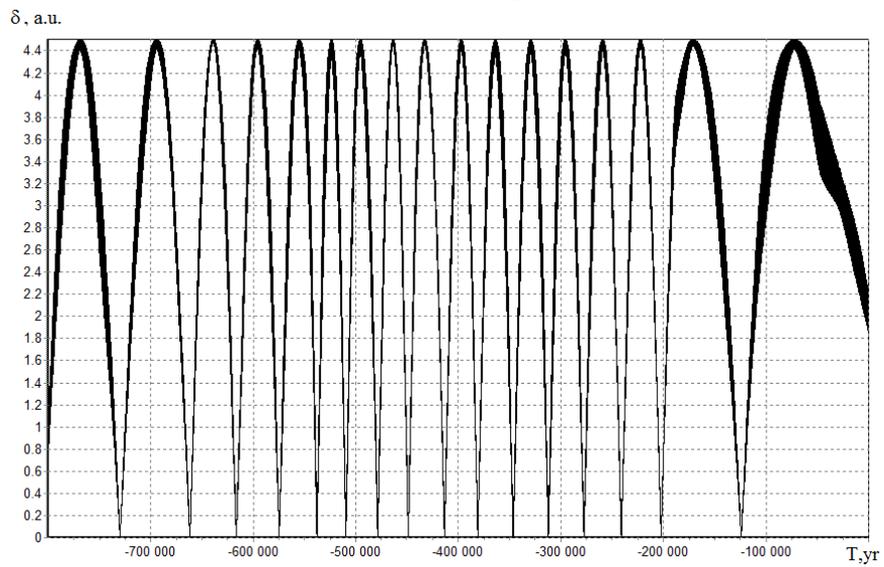

**Fig.12** Evolution distance between 11842 Kap'bos and 228747(2002VH3)

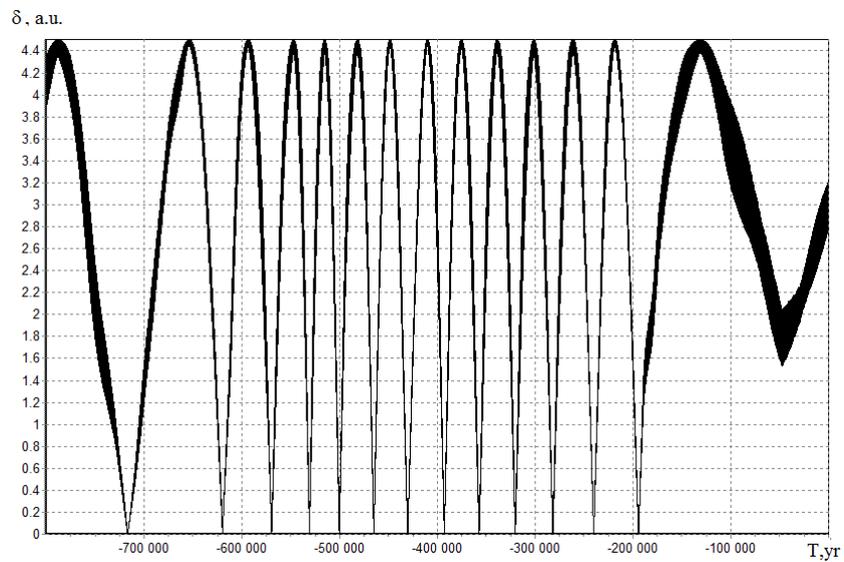

**Fig.13** Evolution distance between 228747(2002VH3) and 436415 (2011 AW46)

Conclusions

We show that resonance-related chaos can play a very important role in dynamics of very young (younger than 1 Myr) asteroid families. The case of high order mean motion resonance (Datura family) is different from case secular resonance (Hobson family). In the high order resonance transfer from initial to final orbit take place by temporary capture in exact resonance. The effect of resonance is very selective in both cases: we see very stable orbits in the vicinity of chaotic ones. The large asteroids (Ceres, Vesta) can made significant effect on dynamic of small bodies in resonance.

On our opinion, the very young compact asteroid clusters are the natural laboratory to study resonance related chaotic and nonlinear dynamics.